\documentstyle[12pt]{article}

\newcommand{\BF}[1]{\mbox{\boldmath $#1$}}

\def\abstracts#1#2#3{{
        \centering{\begin{minipage}{4.62in}\baselineskip=13pt
        \small
        \centerline{\bf Abstract}
        \vspace*{0.2cm}                
        \parindent=0pt #1\par
        \parindent=18pt #2\par
        \parindent=15pt #3
        \end{minipage} }\par}}
\renewcommand{\thefootnote}{\fnsymbol{footnote}}
\begin{document}
\vspace*{-2cm}
\hfill \parbox{4cm}{ FUB-HEP 9/92 \\
               HLRZ Preprint 56/92 \\
               August 1992}\\
\vspace*{2cm}
\centerline{\LARGE \bf Finite-Size Scaling Study of the}\\[0.2cm]
\centerline{\LARGE \bf Three-Dimensional Classical}\\[0.4cm]
\centerline{\LARGE \bf Heisenberg Model\footnotemark}\\[0.3cm]
\footnotetext{\noindent Work supported
in part by Deutsche Forschungsgemeinschaft
under grant Kl256.}
\addtocounter{footnote}{-1}
\renewcommand{\thefootnote}{\arabic{footnote}}
\vspace*{0.2cm}
\centerline{\large {\em Christian Holm\/}$^1$ and
                   {\em Wolfhard Janke\/}$^2$}\\[0.4cm]
\centerline{\large    $^1$ {\small Institut f\"{u}r Theoretische Physik,
                      Freie Universit\"{a}t Berlin}}
\centerline{    {\small Arnimallee 14, D-1000 Berlin 33, Germany}}\\[0.2cm]
\centerline{\large    $^2$ {\small H\"ochstleistungsrechenzentrum,
                      Forschungszentrum J\"ulich}}
\centerline{    {\small Postfach 1913, D-5170 J\"ulich, Germany }}\\[1cm]
\vspace*{2.3cm}
\abstracts{}{We use the single-cluster Monte Carlo update algorithm
to simulate the three-dimensional classical Heisenberg model
in the critical region
on simple cubic lattices of size $L^3$ with $L=12, 16, 20, 24, 32, 40$,
and $48$. By means of finite-size scaling analyses we compute high-precision
estimates of the critical temperature and the critical exponents, using
extensively histogram reweighting and optimization techniques.
Measurements of the
autocorrelation time show the expected reduction of critical slowing down
at the phase transition. This allows simulations on significantly larger
lattices than in previous studies and consequently a better control over
systematic errors in finite-size scaling analyses.
}{}
\thispagestyle{empty}
\newpage
\pagenumbering{arabic}
The critical behaviour of the three-dimensional (3D) classical Heisenberg
model, as one of the simplest spin models, has been investigated
by a variety of approaches. Despite this fact there are still some
discrepancies to be resolved. Motivated by conflicting estimates
of its critical coupling $\beta_c$
on a simple cubic lattice coming from widely accepted high-temperature
series expansion analyses \cite{oldseries} ($\beta_c \approx  0.6916$)
and more recent transfer-matrix (TM) Monte Carlo (MC)
investigations \cite{bloete} ($\beta_c = 0.6922(1)$ and $\beta_c = 0.6925(3)$),
Peczak {\em et al.} \cite{peczak91} (PFL) have recently undertaken a high
statistics
MC study of this model on cubic lattices of sizes up to $V = L^3 = 24^3$.
By simulating the system with the
standard Metropolis algorithm \cite{metropolis} and making extensive use of
multi-histogram techniques \cite{fs}, they could not decide between the two
alternatives. Rather, from a finite-size scaling (FSS) \cite{barber}
analysis of the crossing points of Binder's
cumulant \cite{bindercumu} they claimed an even larger value
of $\beta_c = 0.6929(1)$. Moreover, extrapolations of the
locations of the
susceptibility and specific-heat peak maximum to the infinite volume limit
yielded \cite{peczak91}, respectively,
$\beta_c = 0.6930(2)$ and $\beta_c = 0.6931(10)$, in agreement with the
cumulant crossing value.

However, PFL did not mention a later reanalysis \cite{newseries}
of the high-tem\-per\-a\-ture series expansion based on the Pad\'e
($\beta_c =0.6924(2)$) and ratio ($\beta_c = 0.6925(1)$) method, respectively.
While these values are consistent with the TM estimates given in
ref.\cite{bloete}, we are now faced with the problem that the latest
MC result is significantly higher.

The critical coupling is a non-universal parameter and from this point of view
not of particular interest. Most estimates of universal critical indices,
however, are biased and usually depend quite strongly on the precise value
of $\beta_c$. To clarify the above discrepancy we found it therefore
worthwhile to perform an independent high precision MC study on larger lattices
of sizes up to $48^3$ and with even higher statistics as in PFL's work.
Relying on the Metropolis algorithm as PFL did, such a project would have
been hardly
feasible. As with most local algorithms, the Metropolis (pseudo) dynamics
suffers from the severe problem of critical slowing down, that is from large
autocorrelation times $\tau = a L^z$ (with dynamical critical exponent
$z=1.94(6)$ (ref.\cite{peczak90}) and\footnote{
This can be read off from Fig.~2 in ref.\cite{peczak90}.} $a \approx 3.76$),
which reduce the size of the statistical sample,
$N$, effectively to $N_{\rm eff} = N/2\tau$. It is therefore
crucial to use one of the improved algorithms \cite{improved} of the
past few years that avoid this problem. We chose the cluster
algorithm \cite{sw} in its single-cluster variant \cite{wolff}.
{}From studies of related spin models it is known \cite{othermodels}
that this update algorithm is extremely efficient in three dimensions.

The classical Heisenberg model is defined by the partition function
\begin{equation}
Z = \prod_i \left[ \int
\frac{d\phi_i d\cos\theta_i}{4\pi} \right] e^{-\beta E},
\label{eq:1}
\end{equation}
where $\beta \equiv 1/T$ is the (reduced) inverse temperature
and
\begin{equation}
E = \sum_{\langle i,j \rangle} \left[
1-\vec{s}_i \cdot \vec{s}_j \right]
\label{eq:2}
\end{equation}
is the total energy
(in the following small letters always denote the
corresponding intensive quantities, e.g., $e \equiv E/V$). The sum runs
over all nearest neighbour pairs $\langle i,j \rangle$ and the
three-dimensional unit spins $\vec{s}$ at the sites $i$ of a simple cubic
lattice are parametrized as $\vec{s} = (\cos \phi \sin \theta,
\sin \phi \sin \theta, \cos \theta)$. We always employ periodic boundary
conditions.

Our simulations were organized as follows. First,
we did one run for each lattice size at $\beta_0 = 0.6929$, the estimate of
$\beta_c$ by PFL, and recorded the energy histogram $P_{\beta_0}(E)$ and the
microcanonical averages $\langle\!\langle m^k \rangle\!\rangle(E)
\equiv \sum_M P_{\beta_0}(E,M)
m^k/P_{\beta_0}(E)$,
$k=1,2,4$,
where $m = |\vec{m}|$ is the magnitude of the magnetization $\vec{m} =
\frac{1}{V} \sum_{\BF{x}} \vec{s}(\BF{x})$ of a single spin configuration.
The temperature independent averages
$\langle\!\langle m^k \rangle\!\rangle(E)$
can be computed by accumulating the values of $m^k$ in lists
indexed by the associated energy bin of the configuration and normalizing
at the end by the total number of entries in each bin, making it thus
unnecessary to store the two-dimensional
histogram $P_{\beta_0}(E,M)$.
The continuous energy range $0 \le E \le 3V$ was
discretized into 90000 bins.
The data of this run is sufficient to compute
the approximate
positions $\beta_- < \beta_0$ and $\beta_+ > \beta_0$ of the (connected)
susceptibility and the specific-heat peak maximum
by reweighting techniques \cite{fs} . We then performed two more runs at
$\beta_-$ and $\beta_+$, respectively,
again recording $P_{\beta}(E)$ and $\langle\!\langle m^k \rangle\!\rangle(E)$.
This choice has the advantage that one automatically stays in the critical
region since both $\beta_-$ and $\beta_+$ scale with $L^{-1/\nu}$, where $\nu$
is the correlation length exponent.
{}From this data we can compute three estimates
${\cal O}_L^{(n)}(\beta)$, $n = -,0,+$ for all thermodynamic observables
${\cal O}_L$ of interest,
and for any $\beta$ value in the vicinity of
$\beta_-$, $\beta_0$, $\beta_+$ by reweighting .
Furthermore, since we devided the whole run into several blocks
and stored the energy histograms and microcanonical averages for each block,
we could compute
jackknife errors \cite{jack} $\Delta {\cal O}_L^{(n)}$ on ${\cal O}_L^{(n)}$.
This allowed
us to get an optimized average of these three values that minimizes the
relative error of the combined ${\cal O}_L(\beta)$ for each observable
separately
(the relative weights are simply $1/(\Delta {\cal O}_L^{(n)})^2$).
All our runs contain at least $10000 \times \tau$ measurements, where $\tau$
is the integrated autocorrelation time of the susceptibility. As expected
for the single-cluster update, $\tau$ turns out to be almost independent of the
lattice size and to be very small ($<2$, in units of
lattice sweeps that allow direct comparison with the Metropolis algorithm).
For the $48^3$ lattice, $\tau$ is about
three orders of magnitude smaller than for the Metropolis algorithm. This
explains why we could study much larger lattice sizes than PFL, and could
still afford to have about ten times better statistics.

To determine $\beta_c$ we first concentrated on Binder's cumulant
\cite{bindercumu}
\begin{equation}
U_L(\beta) = 1 - \frac{\langle m^4 \rangle}
{3 \langle m^2 \rangle^2},
\label{eq:3}
\end{equation}
where the angular brackets $\langle \dots \rangle$ denote thermal averages
with respect to (\ref{eq:1}). It is well known \cite{bindercumu} that,
asymptotically for large $L$, all curves $U_L(\beta)$ should cross in the
unique point  $(\beta_c,U^*)$.
Our results  for lattices of size $L$=12, 16, 20, 24, 32, 40, and 48,
obtained by the optimization procedure described above, are shown in Fig.~1.
We see that all curves indeed cross each other at approximately the same
$\beta$ value of $0.693$. A closer look, however, reveals that the crossing
points of $U_L$ and $U_{L'}$ are systematically shifted, depending on the
ratio $b \equiv L/L'$. This is the expected behaviour for finite lattices,
caused by confluent correction terms. Employing well-known \cite{bindercumu}
extrapolation formulas we obtain the final estimates \cite{us}
\begin{equation}
\beta_c =0.6930 \pm 0.0001,
\label{eq:4}
\end{equation}
and
\begin{equation}
U^* = 0.6217 \pm 0.0008.
\label{eq:5}
\end{equation}
The critical coupling is thus found in excellent agreement with the value
quoted by PFL, and also $U^*$ agrees very well with their estimate of
$0.622(1)$. For comparison, a field theoretic expansion in
$\sqrt{4-D}$ predicts for $D=3$ a $4\%$ lower value
of $0.59684...$ \cite{brezin}.

Let us now turn to the FSS estimates of critical exponents.
The derivatives $dU_L/d\beta$ at $\beta_c=0.6930$
should scale asymptotically for large $L$ with $L^{1/\nu}$. In a log-log
plot of all our data points we find a perfect straight line fit (with
goodness-of-fit parameter $Q=0.61$) and from the inverse slope we read off
\begin{equation}
\nu = 0.704 \pm 0.006,
\label{eq:6}
\end{equation}
which again is in agreement with the value quoted by PFL,
$\nu = 0.706(9)$ (determined by the same method, but at $\beta = 0.6929$), and
with the field theoretical estimates of $\nu = 0.705(3)$
(resummed perturbation expansion \cite{pert}), $\nu = 0.710(7)$ (resummed
$\epsilon$-expansion \cite{epsilon}). The high quality of this fit (as well
as of all other fits described below) shows that the asymptotic scaling
formula works down to our smallest lattice size $L=12$, indicating that
there is no need for confluent correction terms.

The ratio of exponents $\beta/\nu$ follows from the scaling of the
magnetization, $\langle m \rangle \propto L^{-\beta/\nu}$.
In a log-log plot of $\langle m \rangle$ at
$\beta_c = 0.6930$ vs $L$ we obtain from a
straight line fit (with $Q=0.68$) the estimate
\begin{equation}
\beta/\nu = 0.514 \pm 0.001,
\label{eq:7}
\end{equation}
which is slightly lower than the value given by PFL, $\beta/\nu = 0.516(3)$
(determined at $\beta=0.6929$). To test by how much our result is biased
by the value of $\beta_c$ we have redone our analysis at $\beta = 0.6929$.
Here we get the slightly higher value of $0.519(1)$. The
quality of the fit, however, is much worse ($Q=0.30$). Since we observe
a similar worsening of the fit at $\beta=0.6931$ ($\beta/\nu = 0.509(1)$,
$Q=0.31$), we take this as support for our estimate of $\beta_c$.
We rely on the goodness-of-fit parameter since visually it
is impossible to make a distinction between these fits when plotted on
a natural scale. It
should be emphasized that even these slight variations in the estimate of
the critical coupling produce  significant changes in the estimate of the
exponent ratio that clearly  dominate the statistical errors. Combining
(\ref{eq:6}) and (\ref{eq:7}) we get for the critical exponent $\beta =
0.362(4)$, to be compared  with the estimates of $\beta = 0.3645(25)$
(ref.\cite{pert}) and $\beta = 0.368(4)$ (ref.\cite{epsilon}).

All other critical exponents follow from (hyper-) scaling relations.
In this way we obtain
\begin{equation}
\alpha/\nu = 2/\nu - D  = -0.159 \pm 0.025,
\label{eq:8}
\end{equation}
\begin{equation}
\gamma/\nu = 3 - 2 \beta/\nu = 1.972 \pm 0.002,
\label{eq:9}
\end{equation}
and
\begin{equation}
\eta = 2 - \gamma/\nu = 2 \beta/\nu - 1 = 0.028 \pm 0.002.
\label{eq:10}
\end{equation}

To test these relations we have performed also a direct analysis of the
critical behaviour of the specific heat
\begin{equation}
C = V\beta^2 \left( \langle e^2 \rangle - \langle e \rangle^2 \right)
\propto L^{\alpha/\nu},
\label{eq:11}
\end{equation}
and of the (connected) susceptibility
\begin{equation}
\chi^{\rm c} = V \beta \left( \langle m^2 \rangle -
\langle m \rangle ^2 \right) \propto L^{\gamma/\nu}.
\label{eq:12}
\end{equation}
Since the magnetization vanishes at and above the critical temperature
we may also take
\begin{equation}
\chi = V \beta  \langle m^2 \rangle \propto L^{\gamma/\nu},
\label{eq:13}
\end{equation}
which is usually a less noisy estimator.

{}From the location of the
maxima of $C$ and $\chi^{\rm c}$ we can get further estimates for
the critical coupling by assuming the FSS relation
$T_{\rm max} = T_c + a L^{-1/\nu} + ...$. Using our value of $\nu = 0.704$
we obtain from the linear fits shown in Fig.~2 the estimates
$\beta_c = 0.6925(9)$ (from $T_{C_{\rm max}}$ with $Q=0.80$) and
$\beta_c = 0.6930(3)$ (from $T_{\chi^{\rm c}_{\rm max}}$ with $Q=1.0$),
respectively. These values are consistent with the cumulant crossing
value (\ref{eq:4}), but have larger statistical errors.

{}From the scaling
of $\chi^{\rm c}$ and $\chi$ at our best estimate of $\beta_c = 0.6930$
we get from linear fits $\eta = 0.0156(44)$ ($Q=0.69$) and
$\eta = 0.0271(17)$ ($Q=0.78$), respectively. The latter value is in
perfect agreement with the scaling prediction (\ref{eq:10}). Moreover,
performing fits to $\chi$ at $\beta = 0.6929$ and $\beta = 0.6931$ we
obtain estimates of $\eta = 0.0364(17)$ ($Q=0.36$) and
$\eta = 0.01781(17)$ ($Q=0.43$), respectively, that are again in perfect
agreement with the scaling predictions based on the corresponding fits
to the magnetization. This is not unexpected since the
measurements of $\langle m \rangle$ and $\chi$ are of course strongly
correlated. Finally, analyzing the FSS behaviour of the susceptibility maximum,
$\chi^{\rm c}_{\rm max} \propto L^{\gamma / \nu}$, we estimate $\eta =
0.0231(61)$ ($Q=0.60$). Notice that all MC estimates are lower than the
field theory values which are $\eta = 0.033(4)$ (ref. \cite{pert}) and $\eta =
0.040(3)$ (ref. \cite{epsilon})\footnote{
In a recent recalculation \cite{recal} of all Feynman graphs several
errors in the
highest order of the $\epsilon$-expansion were corrected. A subsequent
reanalysis \cite{reana} of the resummed series gave a slightly smaller value
for
$\eta$. Compared with the error bar, however, this correction is
negligible.}.

The specific heat exhibits a finite, cusp-like
singularity, because $\alpha$ has a negative
value. We tried a three-parameter fit of the form
$C_{\rm max} = a - b L^{\alpha/\nu}$. The result
$\alpha/\nu = -0.33(22)$ ($Q=0.69$) is compatible with the scaling
relation (\ref{eq:8}) but, due to its very large
statistical error, it does not provide a stringent test of this scaling
prediction. Another way of testing eq.~(\ref{eq:8}) is to assume the
predicted value of $\alpha/\nu = -0.159$ and to fit only the parameters
$a$ and $b$. The resulting fit turned out to be of almost equally good
quality \cite{us}.

In summary, using high-precision data from single-cluster MC simulations
combined with optimized multi-histogram techniques, we have performed
a fairly detailed FSS analysis of the 3D classical Heisenberg model on
simple cubic lattices of size up to $48^3$.
Qualitatively, our main result is that the asymptotic FSS region sets
in for small lattices sizes, $L \approx 12$.
Quantitatively, our value for the critical coupling, $\beta_c = 0.6930(1)$,
is in almost perfect agreement with the MC estimate reported recently
by  PFL (ref.~\cite{peczak91}), but is significantly higher than estimates
from high-temperature series expansion analyses and transfer matrix methods.
Our results for the two basic critical exponents, $\nu = 0.704(6)$ and
$\beta = 0.362(4)$, are in good agreement with field theoretic
predictions. Scaling relations imply $\alpha = -0.112(18)$ and
$\eta = 2 - \gamma/\nu = 0.028(2)$.
Direct measurements of these exponents provide tests of the scaling
relations. In the case of $\eta$ we find good agreement when the
scaling of $\chi$ at $\beta_c$ is considered. Using $\chi^{\rm c}$ at
$\beta_c$ or $\chi^{\rm c}_{\rm max}$, however, the situation is less clear.
In the case of $\alpha$, its negative value causes numerical problems,
since a finite, cusp-like singularity is notoriously difficult to analyze.
%
%
                 \section*{Acknowledgement}
%
The numerical simulations were performed on the CRAY X-MP and Y-MP of
the Konrad-Zuse Zentrum f\"ur Informationstechnik Berlin (ZIB) and
the CRAY X-MP at the Rechenzentrum der Universit\"at Kiel.
We thank both institutions for their generous support.
%
     
\newpage
%
%
  {\Large\bf Figure Headings}
  \vspace{1in}
  \begin{description}
    \item[\tt\bf Fig. 1:]
Fourth-order cumulant $U_L$ vs $\beta$. The values of $U_L(\beta)$ were
obtained by reweighting and optimized combining the results of our
three simulations at different
temperatures for each lattice size $L$.
The simulations were performed at
$\beta_0=0.6929$ (the critical coupling found by
Peczak {\em et al.\/} \cite{peczak91}),
and at the temperature locations of the maxima of the specific heat $C$ and
the susceptibility $\chi^{\rm c}$, respectively.
    \item[\tt\bf Fig. 2:]
Variation of the pseudo transition temperatures $T_{\chi^{\rm c}_{\rm max}}(L)$
and $T_{\rm C_{\rm max}}(L)$  with $L^{-1/\nu}$, where $\nu=0.704(6)$ is our
FSS estimate (see text). The fits yield estimates of $\beta_c = 0.6930(3)$
($Q=1.0$) and
 $\beta_c = 0.6925(9)$ ($Q=0.80$), respectively.
  \end{description}
\end{document}